\documentclass[pre,aps,twocolumn,showpacs]{revtex4}

\usepackage[dvips]{graphicx}
\usepackage{amssymb,amsfonts,amsmath}
\usepackage{color}
\usepackage{ulem}
\usepackage{textcomp}
\usepackage{gensymb}

\newcommand{\rv}{{\mathbf r}}

\newcommand{\J}{{\bf J}}

\newcommand{\F}{{\bf F}}

\newcommand{\vel}{{\bf v}}

\newcommand{\chib}{{\boldsymbol \chi}}

\newcommand{\rnt}{{\bf r}^N\!\!,t}
\newcommand{\kT}{k_{\rm B}T}

\newcommand{\joe}{\color{black}}

\newcommand{\Ca}{{\sf Ca}}

\begin{document}

\author{J.~Reinhardt}
\author{A.~Scacchi}
\author{J.M.~Brader}
\email{joseph.brader@unifr.ch}
\affiliation{Department of Physics, University of Fribourg, CH-1700 Fribourg, Switzerland}

\title{Microrheology close to an equilibrium phase transition}


\begin{abstract}
We investigate the microstructural and microrheological response to a tracer 
particle of a two-dimensional colloidal suspension under thermodynamic conditions close to a liquid-gas 
phase boundary. 
On the liquid side of the binodal, increasing the velocity of the (repulsive) 
tracer leads to the development of a pronounced cavitation bubble, within which the concentration 
of colloidal particles is strongly depleted. 
The tendency of the liquid to cavitate is characterized by a dimensionless ``colloidal cavitation" 
number. 
On the gas side of the binodal, a pulled (attractive) tracer leaves behind it an 
extended trail of colloidal liquid, arising from downstream advection of a wetting 
layer on its surface.     
For both situations the velocity dependent friction is calculated.  
\end{abstract}

\maketitle

\section{Introduction}
The equilibrium bulk phases of colloidal soft matter are now rather well understood, both from a theoretical and 
an experimental perspective.  
However, in many real-world situations the occurrence of stable bulk colloidal 
systems is rather the exception than the rule. 
In practical situations almost all colloidal soft matter is to be found in a state of nonequilibrium, 
undergoing either transient relaxation or driven out of equilibrium by chemical, thermal or mechanical 
perturbation. 

Mechanical driving forces have proven to be particularly useful probes of the structural and 
dynamical properties of soft materials. 
In traditional rheology the perturbing force is applied at the sample boundaries, leading to macroscopic 
deformation of the sample \cite{macosko,larson,mewis,joe_review}. 
The measured relationship between stress and strain provides information about the relevant relaxation 
times of the system and informs the development of constitutive relations \cite{butterworths}. 
However, the microstructural changes underlying the observed rheological response remain unresolved. 
Improved spatial resolution can be achieved by measuring the response to locally applied stress 
or strain fields, implemented by pulling tracer particles through the sample 
\cite{swan,voigtmann_review,squires}.
This method, termed active microrheology, probes the local environment and can be used 
to determine the effective viscosity (``microviscosity") experienced by the tracer \cite{mewis}. 
Active microrheology has traditionally been implemented by measuring the response of dispersed 
magnetic particles to an applied magnetic field \cite{seifriz}. 
More recently, laser tweezers have provided finer control, enabling individual tracer particles to be 
driven along specified paths. 

Regardless of whether the driving forces are applied to a suspension at a global or local level, the richest
physics is often encountered when the thermodynamic state of the quiescent system lies in the
vicinity of an equilibrium phase transition. Within this regime, applied stress or strain fields can interact
strongly with the underlying free energy landscape to produce a nonlinear response, even at relatively low strain rates. 
Particular attention has been given to the question of how macroscopic shear flow can influence the phase 
separation dynamics of an initially homogeneous system which is quenched into the coexistence region 
\cite{bray,dhont2002}. It has been found that during the process of spinodal decomposition under shear 
the morphologies and kinetics which emerge are qualitatively 
different from those observed in the absence of shear \cite{onuki}. 
Phase separating colloidal systems have also been studied in sedimentation experiments
\cite{aarts} and under shear in both experiment 
\cite{derks2008} and theory \cite{stansell}. 

Situations for which external forcing (e.g. gravity or shear) perturbs the
equilibrium phase transition dynamics differ from those which involve phase transitions both initiated and 
sustained solely by the external force field. 
This latter type of drive-induced phase transition are potentially more interesting, as they would not occur in 
the absence of external forces; the
thermodynamic statepoints considered lie outside coexistence regions in the equilibrium phase diagram. 
In polymer solutions, for
example, the phenomena of shear-induced demixing is well known 
\cite{onuki1997,han2006} and for large amplitude
oscillatory shear can lead to the formation of micro-phase separated states with sharp interfacial boundaries
\cite{saito2003}. 
For the case of monodisperse colloidal suspensions, experiments have revealed that crystallization can
be induced in colloidal liquid states close to, but below, the equilibrium freezing transition by applying 
an oscillatory
shear stress \cite{ackerson,besseling2012}. 
These shear-induced crystals thus represent true out-of-equilibrium states; the microstructure relaxes back 
to that of a fluid following cessation of the flow.


In the world of molecular liquids a well known type of local phase transition is cavitation. 
When a solid object moves rapidly through a liquid the local pressure is increased at the 
front of the object and decreased behind it. 
A local phase separation, cavitation, occurs when the pressure behind the object falls below the 
saturated vapour pressure, leading to the development of a gas bubble (called a cavity) in the wake 
\cite{batchelor}. 
This phenomenon, which can have a profound 
effect on the performance of many devices (e.g. ship propellers suffer damage from 
vapour-filled wakes \cite{chen_isrealachvili}), presents a difficult problem for fluid dynamics; one has 
to deal with a nontrivial coupling between the velocity field and thermodynamic phase 
instabilities in the liquid. 
Cavitation remains an active research field with many important applications to pipe flows and 
marine engineering. 

Given the ubiquity of cavitation in molecular liquids it is natural to enquire whether 
similar physics can be observed in colloidal suspensions. 
Although the formal analogy between colloidal and atomic systems (colloids as ``big atoms") 
breaks down in nonequilibrium, qualitatively similar behaviour may still be anticipated in 
situations for which inertia and solvent hydrodynamics are not dominant physical mechanisms. 
Experiments on colloidal cavitation and related phenomena may thus provide insight into 
fundamental physical processes underlying local phase transformation. 
In particular, the increased length- and time-scales presented by colloidal particles 
enable the motion of individual particles to be tracked using confocal microscopy.  
Moreover, tailoring the tracer surface chemistry may be expected to provide a very rich phenomenology.

In this paper we employ a simple model system to investigate 
the microstructural response of a suspension to the motion of a tracer particle moving with 
constant velocity. The thermodynamic statepoints investigated lie close to, but outside, 
the coexistence region of the phase diagram. 
We focus on a minimal two-dimensional model in order to 
identify and characterize the phenomenology without excessive computational 
requirements. 
In contrast to the first order freezing/melting transition, for which dimensionality 
is of fundamental importance, the gas-liquid transition is qualitatively similar in two- 
and three-dimensions (excluding critical phenomena). 
We thus anticipate that our findings will remain qualitatively valid for more realistic 
systems.
The method we employ is dynamical density functional theory (DDFT), which is a reliable 
approach to calculating the dynamics of the one-body density. 
The thermodynamic basis of this approximate theory makes it an ideal tool for 
studying the dynamics of systems close to an equilibrium phase boundary. 
As we will be concerned solely with states at low and intermediate volume fraction, the physics 
of particle cageing and glassy arrest, dominant at high volume fraction, do not play a major role.

The paper is structured as follows: 
In section \ref{theory} we set out the theoretical basis for our investigations: 
In \ref{MicroscopicDynamics} we detail the many-body Brownian dynamics, 
in \ref{ddft} we briefly derive the DDFT, 
in \ref{numerics} we provide details of our numerical algorithms and 
in \ref{model} we describe the specific colloidal model under consideration. 
In \ref{cavitation} we consider statepoints on the liquid 
side of the binodal and investigate colloidal cavitation.  
In section \ref{wetting} we consider the influence of an attractive tracer, for which the wetting 
layer around the particle becomes strongly distorted by the flow. 
Finally, in section \ref{discussion} we discuss our findings and give suggestions for future 
work.

\section{Theory}\label{theory}

\subsection{Colloidal dynamics}\label{MicroscopicDynamics}
Assuming that momentum degrees of freedom equilibrate much faster 
than the particle positions, the dynamics of
colloidal particles can be described on a microscopic level by a
stochastic Langevin equation \cite{gardiner}
\begin{align}
  \gamma \left(\frac{\partial \rv_i(t)}{\partial t} -  \vel^{\rm solv}_i(\rv_i,t)\right) = \F_i(\rnt)  + \chib_i(t),
  \label{EQlangevin}
\end{align}
where $\gamma$ is a friction constant related to the bare diffusion
coefficient, $D_0$, by $\gamma=\kT/D_0$, with $k_{\rm B}$
the Boltzmann constant and $T$ the temperature. 
The solvent velocity is given by $\vel^{\rm solv}_i(\rv_i,t)$.
The random force, $\chib_i(t)$, represents the thermal motion of 
the solvent and is auto-correlated according to 
$\left\langle\chib_i(t)\chib_j(t')\right\rangle 
= 2\kT\gamma\delta_{ij}\delta(t-t'){\bf 1}$.
For a given particle configuration, $\rv^N$, the total force 
due to interactions and external fields is given 
by 
\begin{align}
  \F_i(\rnt) = -\nabla_i  U(\rv^N) + \F_i^{\rm ext}(\rv_i,t),
\label{EQconfig_force}   
\end{align}
where $U(\rv^N)$ is the total interaction potential. 
For simplicity we neglect hydrodynamic interactions, which would arise in a full 
treatment of the solvent flow.

An equivalent description of Brownian
dynamics is provided by the time-dependent configurational probability, 
$\Psi(\rnt)$.  The time evolution of this probability is given exactly
by the many-body continuity equation
\begin{align}
  \frac{\partial \Psi(\rnt)}{\partial t} = 
  -\sum_{i}\nabla_i\cdot\J_i (\rnt),
\label{EQsmol}
\end{align}
where the sum is taken over all particles and where the current of particle $i$ is
given by
\begin{align}\label{EQsmol_current} 
\J_i (\rnt) = \gamma^{-1}\Psi(\rnt) \big[ 
  &\F_i(\rnt) + \gamma\vel^{\rm solv}_i(\rnt) \notag\\
  &\;\;\;\;- \kT\nabla_i \ln \Psi(\rnt)\big]. 
\end{align}
The factor in square brackets appearing
in \eqref{EQsmol_current} is the total force acting on particle~$i$.
Taken together, equations \eqref{EQsmol} and \eqref{EQsmol_current} constitute a many-body
drift-diffusion equation called the Smoluchowski
equation \cite{dhont_book}.

\subsection{Dynamical density functional theory}\label{ddft}
Integrating the Smoluchowski equation over all but 
one of the particle coordinates yields an exact, coarse-grained expression for 
the one-body density profile
\begin{eqnarray}\label{eq:eom_rho1}
\frac{\partial \rho({\bf r}_1,t)}{\partial t} = -\nabla_1\!\cdot {\bf j}(\rv_1,t), 
\end{eqnarray}
where the one-body particle flux is given by 
\begin{align}\label{flux}
{\bf j}(\rv_1,t)&=\gamma^{-1}\rho(\rv,t)\Bigg(
k_B T\,\nabla_1 \ln\rho({\bf r}_1,t)  
+\nabla_1 V_{\rm ext}({\bf r}_1,t) \notag\\
&\hspace*{-0.5cm}-\gamma\vel^{\rm solv}(\rv_1,t)+\int \!d{\rv_2} \frac{\rho^{(2)}({\bf r}_1,{\bf r}_2,t)}{\rho(\rv_1,t)}
\nabla_1 {\joe u (r_{12})}
\Bigg), 
\end{align}
with the equal-time, nonequilibrium two-body density, $\rho^{(2)}({\bf r}_1,{\bf r}_2,t)$, 
and pair potential, $\phi(r_{12}) \equiv \phi(|\rv_1 - \rv_2|)$.  
In equilibrium the integral term in \eqref{flux} obeys exactly the sum rule \cite{evans79}
\begin{eqnarray}\label{sumrule}
\int\!d\rv_2\,  \frac{\rho(\rv_1,\rv_2)}{\rho(\rv_1)}\nabla_1 {\joe u(r_{12})}
= \nabla_1\frac{\delta \mathcal{F}^{\rm \joe exc}[\rho({\bf r}_1)]}{\delta \rho(\rv_1)}, 
\end{eqnarray}
where $\mathcal{F}_{\rm ex}[\,\rho({\bf r})]$ is the excess Helmholtz free energy functional dealing with 
interparticle interactions. 
The notation on the right hand side of \eqref{sumrule} denotes a functional derivative. 
Th excess part is one of three contributions to the total free energy
\begin{align}
  \mathcal{F}[\rho]=\mathcal{F}^{\rm id}[\rho] 
+ \mathcal{F}^{\rm exc}[\rho] +
  \int \!d\rv\, \rho(\rv)V^{\rm ext}(\rv),
\label{helmholtz}  
\end{align}
where the ideal gas free energy is given exactly by
\begin{eqnarray}\label{ideal}
\mathcal{F}^{\rm id}[\,\rho(\rv_1)]=\int d{\bf r}_1\, \rho(\rv_1)[\,\ln(\Lambda^3\rho(\rv_1))-1\,], 
\end{eqnarray}
with thermal wavelength $\Lambda$.

Assuming that \eqref{sumrule} holds in nonequilibrium (an adiabatic approximation on the pair correlations 
\cite{reinhardtbrader}), substitution into \eqref{flux} yields the 
DDFT for the one-body density
\begin{align}
\frac{\partial \rho(\rv_1,t)}{\partial t} = \nabla_1\cdot\Bigg[ 
&-\rho(\rv,t)\vel^{\rm solv}(\rv_1,t) \notag\\
&+\gamma^{-1} \rho(\rv_1,t) \nabla_1\frac{\delta \mathcal{F}[\rho(\rv_1,t)]}{\delta \rho(\rv_1,t)} \Bigg].
\label{stDDFT}
\end{align} 
From Eqs.\eqref{flux} and \eqref{sumrule} the DDFT approximation to the current is found to be
\begin{align}
{\bf j}(\rv_1,t)&=-\frac{\rho(\rv,t)}{\gamma}\Bigg(
k_B T\,\nabla_1 \ln\rho({\bf r}_1,t)  
+\nabla_1 V_{\rm ext}({\bf r}_1,t) \notag\\
&\;\;\;\;\;\;\;\;\;\;-\gamma\vel^{\rm solv}(\rv_1,t) + 
\nabla_1\frac{\delta \mathcal{F}^{\rm\joe exc}[\rho({\bf r}_1)]}{\delta \rho(\rv_1)}
\Bigg), 
\label{DDFT_current}
\end{align}
from which is it apparent that the DDFT approximation replaces the real interaction forces in the system 
by an effective one-body force generated from a chemical potential gradient.

\begin{figure}
\begin{center}
\includegraphics[width=8.3cm]{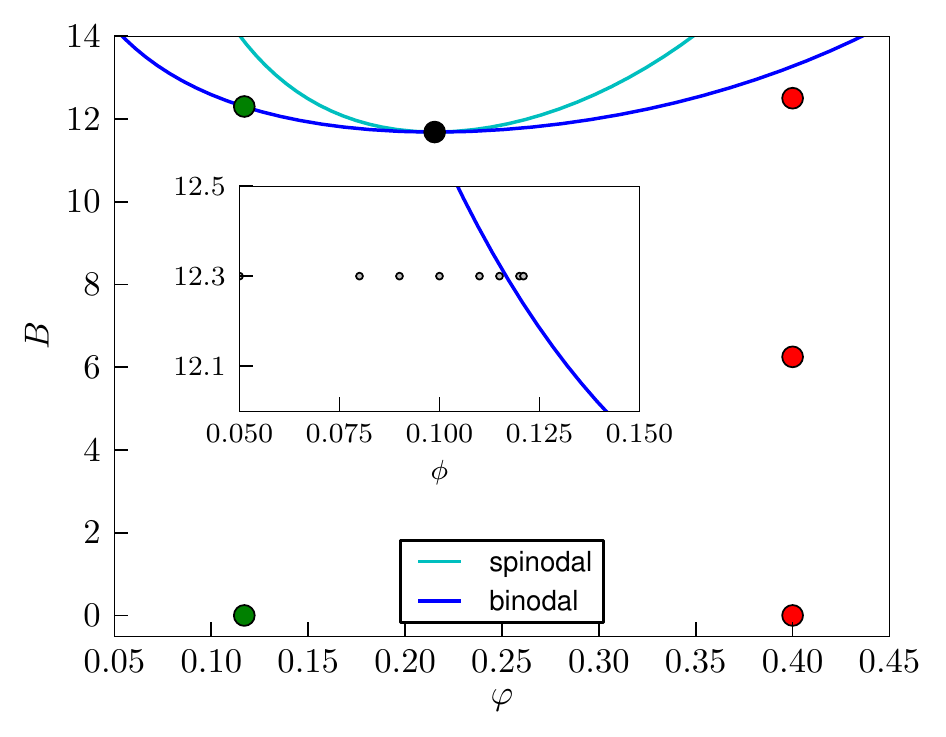}
\end{center}
\caption{\label{figure1}
Phase diagram of the hard-disk attractive square well system. 
$B$ characterizes the integrated strength of the attractive interaction. 
The critical point is at $\phi_{\rm crit}=0.215$ and $B_{\rm crit}=11.68$.
Red and green symbols indicate the thermodynamic statepoints for which we perform detailed 
structural investigations in sections \ref{cavitation} and \ref{wetting}, respectively. 
Inset: Close-up of the binodal (blue line). The points indicate the path taken for 
the calculations in \ref{wetting}. The rightmost point ($\phi=0.121$) lies approximately 
on the numerical binodal.}
\end{figure}

\subsection{Approximate density functional}\label{approximation}
The standard density functional treatment of attractive forces is to separate the excess free energy 
into the sum of a reference free energy and a peturbative mean-field term \cite{evans92}
\begin{align}
  \mathcal{F}^{\rm exc}[\rho]=\mathcal{F}^{\rm exc}_{\rm hd}[\rho] 
+ \frac{1}{2}\int\! d\rv_1\!\! \int\! d\rv_2 \,\rho(\rv_1)\rho(\rv_2)u_{\rm att}(r_{12}), 
\label{meanfield}  
\end{align}
where $u_{\rm att}(r_{12})$ is the attractive component of the total pair potential (defined {\joe here} to be 
zero within the hard core region).

The first term in \eqref{meanfield} is the free energy of a reference system, taken to be a system 
of repulsive hard-disks.
We approximate this reference free energy using a recently
proposed ``fundamental measures" {\joe functional} due to Roth {\it et al.} \cite{oettel}. 
This approximation incorporates correctly the geometry and packing physics of the disk{\joe s}  
(for details see \cite{oettel}). 
The second, mean-field term in \eqref{meanfield} has been applied to a huge range of problems 
in equilibrium liquid state theory and generally provides qualitatively {\joe robust} predictions 
\cite{barker}). 

\subsection{Numerical methods}\label{numerics}

The numerical solution of \eqref{eq:eom_rho1}  and \eqref{DDFT_current} is 
approached in a fashion similar to \cite{reinhardt2013a}, where a test particle method 
was employed to calculate the distorted pair correlation function under shear flow. 
A primary difficulty is that the circular shape of the tracer particle does 
not fit naturally with discretisations based on a cartesian grid. This is
especially true for the region close to the boundary of the excluded volume,
where the numerically determined density is very sensitive to the spatial
discretisation. 
For this reason we employ methods similar to the Finite Element Method (FEM),
which provide a great deal of flexibility for the spatial discretisation. This
makes it possible to choose a very fine grid to properly represent the shape of
the tracer and achieve a high accuracy in critical parts of the computational
domain. For regions far away from the tracer, where the density 
is slowly varing, a coarse grid is sufficient. 
This keeps the computational demand low and allows the treatment
of sufficiently large domains, which is necessary to accomodate the extended
wake structures which develop at higher tracer velocities.

The main complication in the solution of the density functional equations is the
necessity to evaluate nonlocal terms stemming from the excess free energy
contribution \eqref{meanfield}. The Fast-Fourier-Transform techniques which allow for
simple and fast evaluation of convolution integrals on cartesian grids are not applicable 
to more general meshes \cite{roland_review}. As the weight functions appearing in the mean-field and reference parts 
of the excess free energy are independent of the density, the neccessary convolutions 
can be pre-calculated and evaluation of the nonlocal terms reduces to a number of
matrix-vector products.  Due to the finite range of the weight functions, the
pre-calculated convolution matrices have many zero entries, which keeps the memory
requirements manageable. 
As the methods we employ are closely related to the FEM, we can exploit the high-quality 
libraries, tools and utilities which are already available. 
Our custom solvers are based on the finite element framework deal.II \cite{Bangerth2007} 
and the meshes used for our calculations were created using Gmsh \cite{Geuzaine2009}.

\subsection{Model system and parameters}\label{model}

The specific model we employ is a two-dimensional {\joe system of} Brownian hard-disks with a square-well attraction. 
The pair potential for this model is given by 
\begin{align}\label{pair}
u(r) = \left\{
  \begin{array}{lr}
    \infty  \hspace*{-2cm} &  0 < r < \sigma\;\,\\
    -\epsilon \;&  \sigma \le r < \sigma\delta\\
    0  &    r \ge \sigma\delta
  \end{array}
\right.
\end{align} 
where $\epsilon$ is the depth of the attractive well, $\delta$ sets the range and 
$\sigma\equiv 2R$ is the diameter of a disk. 
We will henceforth measure all distances in units of $R$.

As only relative velocities are of physical {\joe significance}, the motion of the 
tracer is implemented by setting the solvent velocity field equal to a constant, 
$\vel^{\rm solv}=-v^{\rm solv}\hat{\bf e}_{\rm x}$, modelling tracer 
motion in the positive $x$-direction. 
We {\joe will henceforth use a dimensionless tracer velocity, defined according to}
\begin{align}\label{bare_peclet}
v \equiv\frac{v^{\rm solv}R}{D_0}, 
\end{align} 
which compares the timescale of external driving, $\tau_{\rm drive}\equiv R/v_{\rm solv}$, 
with that of diffusive relaxation, $\tau_{\rm diff}\equiv R^2/D_0$. 
For values of $v$ greater than unity external driving will dominate diffusive relaxation.

The square-well system is characterized by two independent parameters, $\epsilon$ and $\delta$. 
In the following we will fix the range of the potential to the value $\delta=R$ and report 
our results in terms of the attraction strength parameter
\begin{align}\label{B}
B = -4(\delta^2-1)\epsilon\phi,
\end{align}
where the two-dimensional area fraction is given by $\phi=\pi N/V$, with 
$N$ the number of particles and $V$ the system volume.   
{\joe The parameter $B$ emerges naturally from the bulk limit of the mean-field density functional 
\eqref{meanfield} and enables the phase behaviour of the system for all values 
of $\epsilon$ and $\delta$ to be captured by a single phase diagram in the $(\phi\,,B)$ plane. }

For sufficiently high attraction strength ($B>B_{\rm crit}\approx 11.68$) the bulk free energy, obtained by 
setting $\rho(\rv)=\rho_{\rm b}$ 
in \eqref{ideal} and \eqref{meanfield}, exhibits a van der Waals loop, indicating the onset of gas-liquid phase 
separation. 
In Fig.\ref{figure1} we show the phase diagram of the square well system, including both 
the binodal line enclosing the coexistence region and the spinodal line which marks the boundary of 
mechanical stability.  
The symbols locate the thermodynamic statepoints for which we conduct detailed microstructural 
investigations. 
The inset shows a close-up of the gas side of the binodal and gives the thermodynamic path followed 
for our investigation of wetting in section \ref{wetting}.

\section{Results}\label{results}

\subsection{Cavitation}\label{cavitation}

The first situation we consider is the response of the square-well host suspension to a purely repulsive tracer. 
The total system, consisting of tracer plus suspension, is then fully specified by \eqref{pair} 
together with the tracer-colloid interaction potential, given by
\begin{align}\label{dry_potential}
u_{\rm ct}(r) = \left\{
  \begin{array}{lr}
    \infty &\;\; 0 < r < (R_{\rm t} + R)\\
    0  &\;\;  r \ge (R_{\rm t} + R)
  \end{array}
\right.
\end{align}
where $R_{\rm t}$ is the radius of the tracer disk. 
For our study of cavitation we choose $R_{\rm t}=4R$. 
This value is sufficiently large to clearly distinguish the tracer from 
the colloidal particles, while avoiding unnecessary numerical effort 
associated with very large tracers.

\begin{figure*}
\begin{center}
\hspace*{-0.2cm}\includegraphics[width=16.7cm]{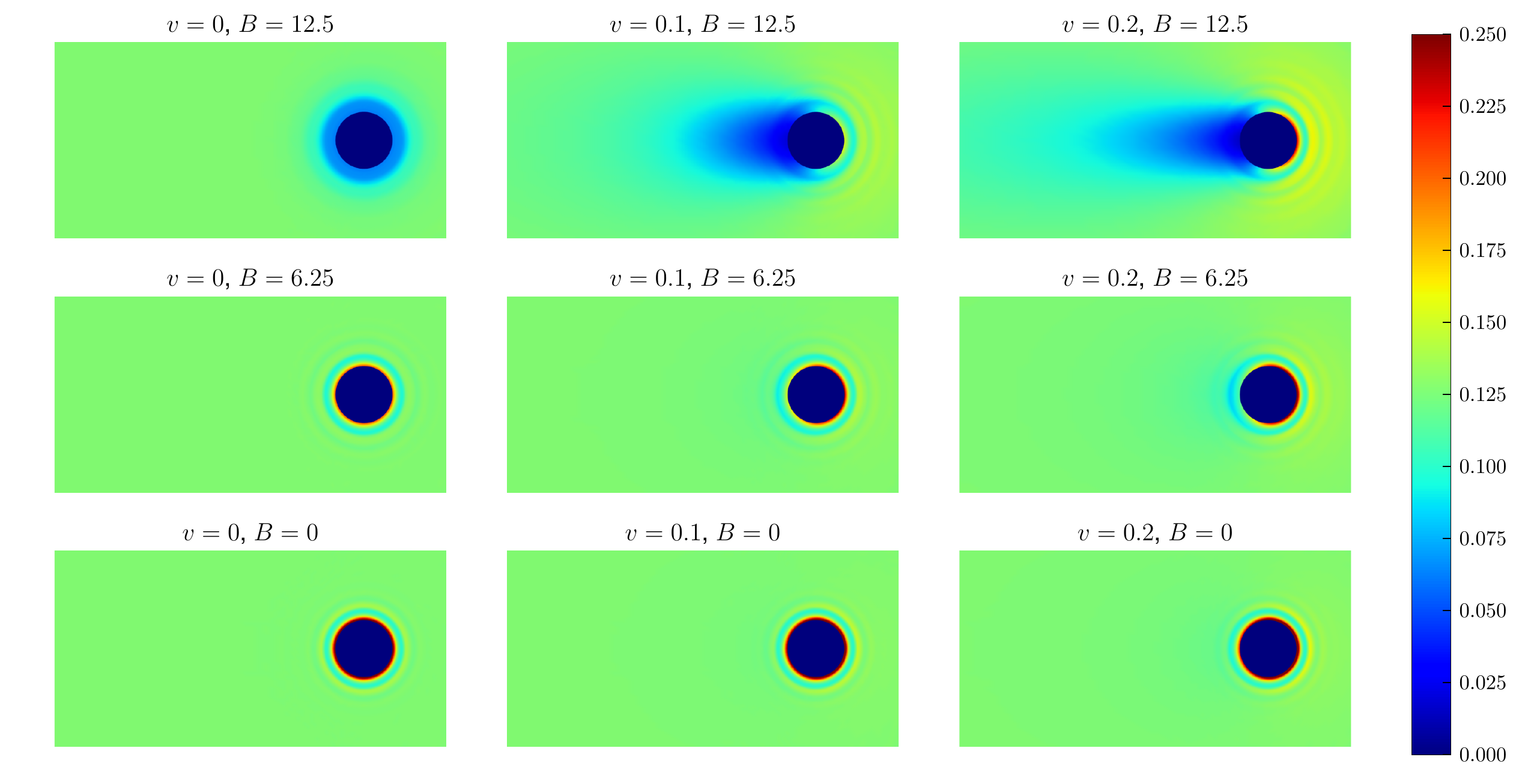}
\end{center}
\caption{\label{figure2}
Density profiles around a tracer pulled with constant velocity. 
The bulk volume fraction is $\phi=0.4$ and the tracer radius $R_{\rm t}=4R$. 
Bottom row: Profiles for pure hard disks ($B=0$) and dimensionless tracer velocities 
(left to right) $v=0\,,\, 0.1\,,\, 0.2$. Only very slight anisotropy develops at finite 
tracer velocity. 
Middle row: Profiles for intermediate attraction strength ($B=6.25$) at the same three values 
of $v$. 
Top row: Profiles for a statepoint close to the binodal ($B=12.5$) illustrating the phenomenon 
of cavitation. Due to the proximity of the binodal, the equilibrium profile ($v=0\,,\,B=12.5$) 
exhibits a drying layer about the tracer surface. 
}
\end{figure*}

\subsubsection{Equilibrium profile}
In the left column of Fig.\ref{figure2} we show two-dimensional density profiles 
about a stationary tracer. 
As the attraction strength is increased for fixed volume fraction, $\phi=0.4$,  
(statepoints indicated in Fig.\ref{figure1}) 
there is relatively little change in the overall density profile, except in the vicinity 
of the contact peak. However, close to the binodal we observe the development of a region 
of gas density, a drying layer, around the tracer surface. 
The phenomenon of drying (``wetting by gas") is well known for 
systems of attractive particles at planar hard walls: the density profile loses its 
oscillatory character and a layer of gas develops at the substrate. 
As the distance from coexistence vanishes the thickness of the gas layer diverges. 
This is a consequence of the fact that the liquid particles are attracted more strongly 
to each other than to the wall. 
Planar drying was first observed in computer simulations 
\cite{abraham,sullivan} and later using a mean field DFT similar to that employed here 
\cite{tarazona} (see also \cite{henderson}).

The behaviour of the drying layer at curved substrates is complicated by the (geometrical) fact 
that the area of the interface between gas and liquid phases changes with the layer thickness. 
Due to the surface tension between gas and liquid, which acts against the increase of interfacial 
area, thick gas layers become energetically unfavorable (for detailed studies in this direction see 
\cite{gelfand,upton,bieker}). 
The drying layer around our circular tracer thus remains finite all the way up to coexistence. 
The top left panel of Fig.\ref{figure2} shows the profile for a statepoint very close to 
coexistence. The drying layer extends to approximately 3R from the tracer surface (see the profile 
indicated by the arrow in Fig.\ref{figure3} for a radial slice).     

\subsubsection{Cavitation profiles}
\begin{figure}
\begin{center}
\includegraphics[width=8.3cm]{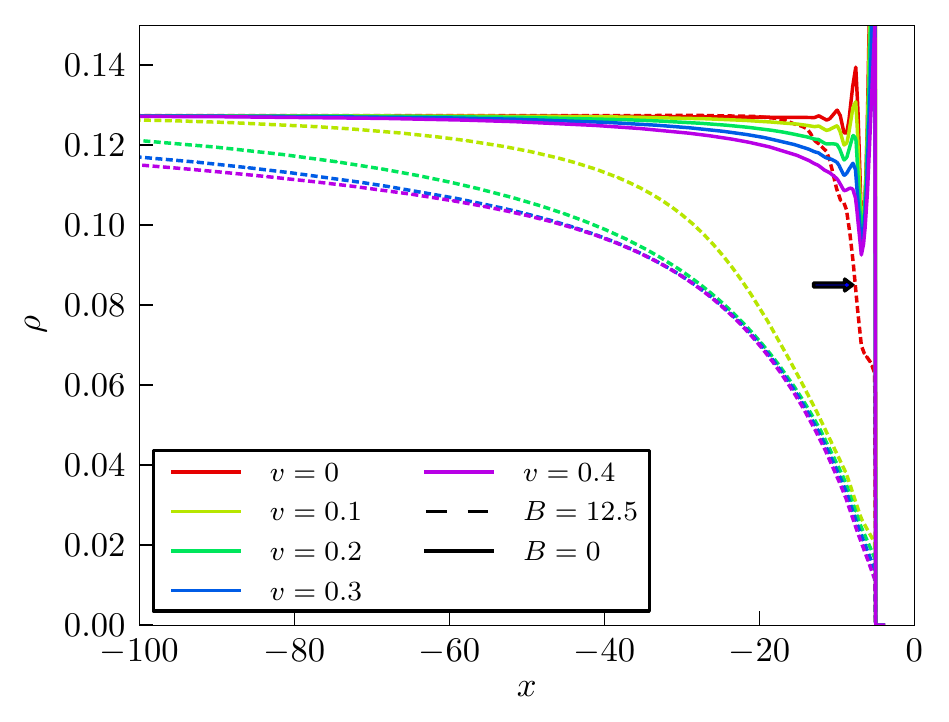}
\end{center}
\caption{\label{figure3}
Focusing on the wake behind the tracer. 
The coordinate $x$ measures the distance from the center of the tracer out through the wake behind it. 
The bulk volume fraction is $\phi=0.4$. Here we compare the wake which develops for statepoints close 
to the binodal (full lines) with that for the pure hard disk system ($B=0$, broken lines). 
For $B=12.5$ and $v>0.4$ the profile saturates and becomes velocity independent in the region $-40<x<0$.}
\end{figure}

In the bottom row of Fig.\ref{figure2} we show the steady state density profile 
for pure hard disks ($B=0$) at velocities $v=0.1$ and $0.2$. 
Although some anisotropy is induced by the tracer motion, there is no 
major disruption of the equilibrium microstructure, as demonstrated by the one dimensional 
slices shown in Fig.\ref{figure3}. 
For the velocities shown there is only a slight reduction in density behind the tracer. 
We find that only for $v>0.8$ does the structure behind the particle become 
significantly modified by the flow (the height of the contact peak behind the 
particle drops below the bulk density). 
This indicates that the reduced velocity \eqref{bare_peclet} provides an appropriate scaling for 
the purely repulsive, hard disk system at the densities considered; microstructural distortion arises from a balance 
between the rate at which colloids being pushed out of the way by the tracer and single particle 
diffusive relaxation to equilibrium. 
For larger area fractions, approaching the glass transition \cite{bayer,hajnal} 
a more appropriate scaling would be provided by the dimensionless velocity $\vel_{\rm solv}\tau_{\alpha}/R$, 
where $\tau_{\alpha}$ is the timescale of structural relaxation. 
A treatment of structural relaxation which incorporates glassy physics (see e.g. \cite{gazuz}) 
lies beyond the scope of adiabatic DDFT \cite{power,noz}.

In the top row of Fig.\ref{figure2} we again show steady state density profiles for different values of $v$, 
but now for a statepoint close to the binodal.
As the velocity is increased the drying layer begins to extend into the region behind the particle, forming 
the colloidal analogue of a cavitation bubble. 
A significant region of colloid depletion can be observed already by velocity $v=0.1$ (top center panel), 
for which the corresponding pure hard disk density profile is barely distorted from its equilibrium form 
(lower centre panel). 
As the velocity is increased to $v=0.2$ the cavitation bubble becomes more pronounced, extending beyond 
$60R$ behind the tracer surface.   
The dimensionless velocity defined in \eqref{bare_peclet} does not capture the correct 
competition of timescales within the ``cavitation regime" of system parameters and it 
is thus desirable to find a more appropriate scaling. 

In Fig.\ref{figure3} we show one-dimensional slices through the density distribution, focusing on the cavitation 
region directly behind the tracer. At equal volume fraction and velocity the cavitating profiles ($B=12.5$) 
differ greatly from those of pure hard disks ($B=0$) and clearly exhibit a large region of reduced colloid 
density. The calculation of these profiles is very time consuming and many iterations 
are required to obtain satisfactory convergence. The numerical challenge here is very similar to that presented by 
standard equilibrium calculations of planar wetting or drying, for which the film thickness only evolves very 
slowly as the solution is iterated. 
We find that roughly within the range $-40<x<0$ the density profile saturates and ceases to change for velocity values 
in excess of $0.4$. 
Increasing the velocity beyond this value leads to a development of the cavitation bubble 
for $x>40$, but no longer modifies the structure closer to the tracer. 
It thus appears that there is some qualitative change in the evolution of the profile with $v$ beyond 
a value around $0.4$. 
Whether the bubble grows indefinitely with increasing $v$ remains unclear, due to 
limitations on the numerical grid sizes we can implement.

\subsubsection{Colloidal cavitation number}
For molecular liquids the parameter which describes the conditions for cavitation in 
steady state flow is the dimensionless ``cavitation number" \cite{batchelor}
\begin{align}\label{cavitation_number}
K \equiv \frac{p_{\rm bulk}-p_{\rm coex}}{\frac{1}{2}\rho_{\rm m} U^2}.  
\end{align}
where $p_{\rm bulk}$ and $p_{\rm coex}$ are the bulk and vapour pressures, respectively, 
$\rho_{\rm m}$ is the mass density and $U$ is the flow velocity at infinity (far from the tracer). 
For a given temperature 
the tendency of a molecular liquid to cavitate at a particular spatial location depends upon the 
the difference between the local pressure, $p(\rv)$, and the vapour pressure, $p_{\rm coex}$. 
The former is not known {\it a priori} and 
depends nontrivially on both the flow and details of the tracer shape. 
In order to obtain a simple constant characterizing the state of the system one can replace the local 
pressure by that of the bulk ({\joe i.e. the }system in the absence of a tracer), which yields the numerator 
in \eqref{cavitation_number}. 
This pressure difference, which has units of energy density, is then compared with the kinetic energy 
density of the liquid, the denominator in \eqref{cavitation_number}, to generate a dimensionless ratio 
$K$.  
It has been found empirically that for values of $K$ below a critical value, typically of order unity, 
the liquid will begin to cavitate. 
The cavitation number thus provides the basis for scaling cavitation phenomena and for designing model experiments 
\cite{streeter}.




In colloidal suspensions inertial effects are not important. The kinetic energy will therefore 
be of no relevance for a scaling argument. However, a natural candidate to replace the kinetic  
energy in \eqref{cavitation_number} is provided by the rate of viscous dissipation in the flowing 
system. 
Within the adiabatic DDFT approximation employed here the local energy loss per unit volume 
per unit time, the dissipated power density, is given by, $\gamma\J^2(\rv)/\rho(\rv)$ \cite{power}.
In order to provide a simple constant characterizing the system we replace spatially dependent 
functions with their bulk values, $\gamma\J^2(\rv)/\rho(\rv)\rightarrow \gamma\rho\vel_{\rm sol}^2$, 
and compare this with the osmotic pressure deviation, $\Pi_{\rm bulk}-\Pi_{\rm coex}$. 
{\joe Dividing the osmotic pressure difference by the power density} 
yields a characteristic timescale for cavitation  
\begin{align}
\tau_{\rm cav} \equiv \frac{\beta(\Pi_{\rm bulk}-\Pi_{\rm coex})}{\rho_{\rm bulk}v^2}\frac{R^2}{D_0},
\end{align}
where we here explicitly include factors of $R$ and where the dimensionless $v$ is given by \eqref{bare_peclet}.  
The ratio of $\tau_{\rm cav}$ to the diffusive relaxation time $\tau_{\rm diff}$ defines a 
dimensionless colloidal cavitation number
\begin{align}\label{colloidal_cavitation_number}
\Ca \equiv\frac{\tau_{\rm cav} }{\tau_{\rm diff} } = \frac{\beta(\Pi_{\rm bulk}-\Pi_{\rm coex})}{\rho_{\rm bulk}v^2}.
\end{align}
This can be neatly rewritten in terms of the compressibility factor, $Z\equiv \beta\Pi/\rho$, as follows
\begin{align}\label{colloidal_cavitation_number2}
\Ca  = \frac{1}{v^2}\left( Z_{\rm bulk} - \frac{\phi_{\rm coex}}{\phi_{\rm bulk}}Z_{\rm coex} \right).
\end{align}
The colloidal cavitation number is a function of the external control parameters $\phi_{\rm bulk}$, $T$ and $v$ 
(where the temperature dependence is obtained via the function $\phi_{\rm coex}(T)$). 
 
For our approximate density functional \eqref{meanfield} the compressibility factor is given by a simple 
function
\begin{align}\label{compressibility_factor}
Z(\phi,B) = \frac{1}{(1-\phi)^2} - \frac{1}{2}B\phi, 
\end{align}
where $B$ is given by \eqref{B}. 
For $B=12.5$ the coexistence volume fraction is given by $\phi_{\rm coex}=0.349$. 
Taking the value $\phi_{\rm bulk}=0.4$ thus yields the relationship between cavitation 
number and velocity ${\sf Ca}= 0.122/v^2$.
If we now assume that a value of ${\sf Ca}$ equal to unity characterizes the onset of cavitation, then 
we obtain a cavitation velocity, $v_{\rm cav}=0.35$. 
This estimate corresponds surprisingly well to the velocity value found in our numerical calculations, 
beyond which the density profile partially saturates (within the range $-40<x<0$) as the velocity is increased.


\begin{figure}
\begin{center}
\vspace*{-0.3cm}
\includegraphics[width=8.2cm]{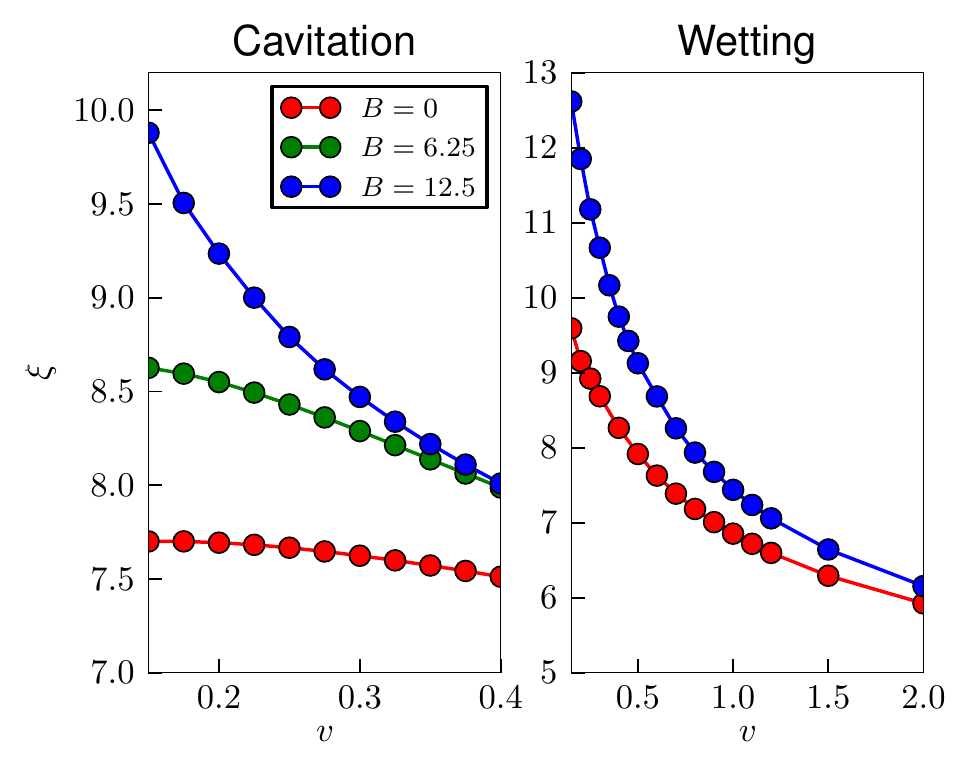}
\end{center}
\caption{\label{figure4}Friction coefficient, $\xi$, as a function of velocity. 
Left panel: Purely repulsive tracer, $\phi=0.4$ for three values of $B$. 
For $B=12.5$ the onset of cavitation leads to a strong reduction 
in the friction as a function of velocity.  
Right panel: Attractive tracer (see Eq.\eqref{wet_potential}) at $\phi=0.117$. 
The friction reduces rapidly as a function of $v$ due to flow induced modification 
of the wetting layer behind the tracer. 
}
\end{figure}

\subsubsection{Friction coefficient}
We define the velocity dependent friction coefficient {\joe as the magnitude ratio of 
the frictional force acting on the particle to the far field velocity} 
\begin{align}\label{friction}
\xi = \frac{|{\bf F}_t|}{|\vel_{\rm solv}|}, 
\end{align}
which is a function of $B$, $\phi$ and $v$.
The force on the tracer is calculated from the density distribution by integration
\begin{align}\label{force_integral}
{\bf F}_t = \int \!d\rv \, \rho(\rv)\nabla V_{\rm ext} (\rv),   	
\end{align}
which acts in the negative $x$-direction. 
The friction defined by \eqref{friction} is related to the microviscosity according to 
$\eta_{\mu}\equiv 6\pi\xi R_t$ \cite{mewis}. 

In the left panel of Fig.\ref{figure4} we show the friction coefficient as a function of 
$v$, for fixed volume fraction $\phi=0.4$. 
For a given velocity, increasing the amount of cohesion between colloids makes it more difficult for the tracer 
to push through them, leading to an increase of $\xi$ as a function of attraction strength. 
For both $B=0$ and $B=6.25$ the friction exhibits a plateau at low velocities, followed by a gradual reduction 
as the velocity is increased. 
For $B=12.5$, however, the friction coefficient shows a stronger reduction with increasing velocity, 
reflecting the onset of cavitation. 
Although we anticipate the emergence of a plateau for small velocities, resolving the friction coefficient 
in this regime is unfortunately beyond the capabilities of our current algorithms. 
At low velocity the friction coefficient \eqref{friction} is formed by the ratio of two 
small numbers, thus requiring high accuracy calculations of the force integral \eqref{force_integral}. 

The velocity dependence of the friction coefficient shown in Fig.\ref{figure4} is reminiscent of, and is 
indeed closely related to, the phenomenon of shear thinning in macroscopic bulk rheology. 
In both cases, changes in the equilibrium configurational distribution enable the flow 
to proceed with a reduced number of particle collisions, and hence a reduced dissipation. 
In bulk systems these configurational changes can take the form of spatial ordering 
(``layering" \cite{joe_review,brader_kruger}). 
In the present case, the packing structure that develops around the front of the tracer, and which moves along with 
it for a certain distance before rolling off, serves to lubricate the motion 
of the tracer through the suspension.    
The formation of a cavity modifies the microstructure around the sides of the tracer 
in such a way that the lubrication effect is enhanced.

\subsection{Wetting}\label{wetting}
\begin{figure}
\begin{center}
\includegraphics[width=8.2cm]{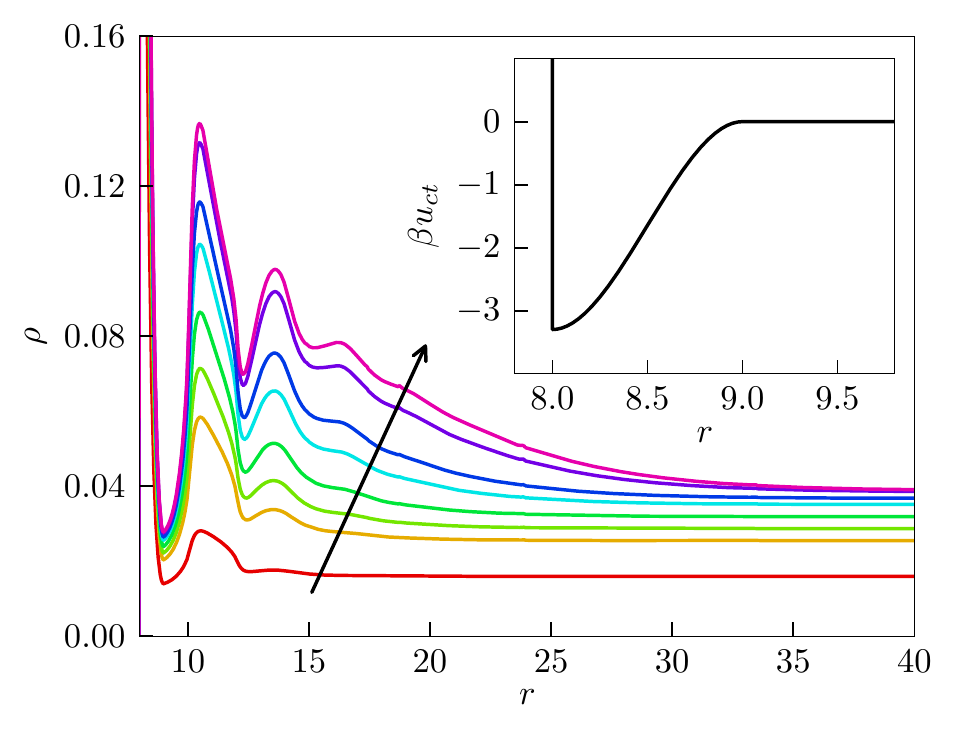}
\end{center}
\caption{\label{figure5}
Equilibrium density profiles at the surface of the tracer particle for 
$B=12.3$ and values of $\phi$ along the path shown in the inset to 
Fig.\ref{figure1}. Profiles are shown for 
$\phi=0.05, 0.08, 0.09, 0.10, 0.11, 0.115, 0.12$ and $0.121$, 
where the arrow indicates the direction of increasing $\phi$. 
A wetting layer around the tracer is clearly visible. 
Inset: The tracer-colloid interaction potential \eqref{wet_potential} for the parameters emplyed in our 
numerical calculations. 
}
\end{figure}

We now consider the more complex situation for which the tracer has an attractive interaction 
with the surrounding colloidal particles.  
The system is thus specified by (13), together with the following tracer-colloid interaction potential
\begin{align}\label{wet_potential}
u_{\rm ct}(r) = \left\{
  \begin{array}{lr}
    \infty            \hspace*{3.15cm}0\le r < R_{\rm tc}\\
    ar^3 + br^2 + cr + d  \hspace*{0.38cm}R_{\rm tc} \le r \le \delta_t R_{\rm tc}\\
    0 \hspace*{3.97cm} r >\delta_{\rm t} R_{\rm tc}\\
  \end{array}
\right.
\end{align}
where $R_{\rm tc}=R_{\rm t} + R$ is the sum of tracer and colloid radii and $\delta_{\rm t}$ sets 
the range of the interaction. 
The particular choice of polynomial form was chosen to faciliate numerical solution of our equations.  
The coefficients, $a,\ldots,d$, were determined by requiring that $u_{\rm ct}(r)$ has a vanishing derivative at 
$r=R_{\rm tc}$ and $r=\delta_{\rm t}R_{\rm tc}$, such that the force entering the DDFT equation 
\eqref{DDFT_current} remains continuous. 
We henceforth employ the values $a=-6.6$, $b=168.3$, $c=-1425.6$ and $d=4009.5$, which correspond to a potential 
depth of $3.3k_{\rm B}T$ and a range of one colloid radius (the potential is shown in the inset to Fig.\ref{figure5}). 
To study the effect of wetting on the tracer surface we choose $R_{\rm t}=7$, which represents a compromise between 
the desired low surface curvature (required to obtain a good size wetting film) and our numerical constraints.  

\subsubsection{Equilibrium profiles}
The physical adsorption of an atomic or molecular gas onto a planar solid substrate is a problem of three-phase 
coexistence and presents a rich phenomenology of surface phase behaviour. 
In addition to the temperature and density, the tendency of a condensed liquid 
droplet to spread out and ``wet" the substrate is governed by the strength of the intermolecular potential 
relative to that exerted on the molecules by the substrate. 
As thermodynamic control parameters are (adiabatically) tuned towards their values on the gas side of the 
binodal, the substrate can become either partially or completely wet by the liquid phase, depending upon 
whether the temperature is above or below the wetting temperature. 
Complete wetting can occur either continously or in a sequence of discontinuous layering transitions, 
for which the adsorption jumps by a finite amount \cite{tarazona_evans,dietrich}. 
  
Recently, it has been shown that the depletion attraction induced (e.g. by adding nonadsorbing polymer) 
between colloids and a substrate can also lead to interesting wetting phenomena 
\cite{wet1,wet3,roth_brader_schmidt,wet2,col_pol_review}. 
Although much of the interface phenomenology is closely analogous to that found in molecular systems, novel surface 
transitions arising from many-body depletion interactions have been predicted by both theory and simulation 
\cite{wet2,col_pol_review}.
Moreover, the surface tension between coexisting colloidal gas and liquid phases is orders of magnitude lower 
than in molecular liquids (of order $\mu N/m$, rather than $N/m$) \cite{squaregradient,els}. 
This is a result of the fact that the surface 
tension scales as $k_{\rm B}T/R^2$, where the colloid radius is typically larger than that of a molecule 
by a factor of $10^3$. 
In a slightly more realistic treatment of colloidal active microrheology our chosen potentials \eqref{pair} and 
\eqref{wet_potential} could be replaced by the Asakura-Oosawa depletion potential depletion potential.

In the present application to microrheology the substrate presented by the tracer surface has a finite curvature. 
Equilibrium wetting at curved or structured substrates is more complicated than the standard planar case and remains an active 
field of research. Most relevant for our present work are the 
publications of Gil and coworkers \cite{gil1,gil2,gil3} who considered theoretically the wetting of circular substrates 
immersed in two dimensional fluids close to phase separation. 
The salient finding is that the positive curvature of the interface prohibits the development of an infinitely thick film of 
liquid as the binodal is approached from the gas side. 
The adsorption at coexistence is thus finite.  

\begin{figure}
\begin{center}
\hspace*{-0.5cm}\includegraphics[width=9.4cm]{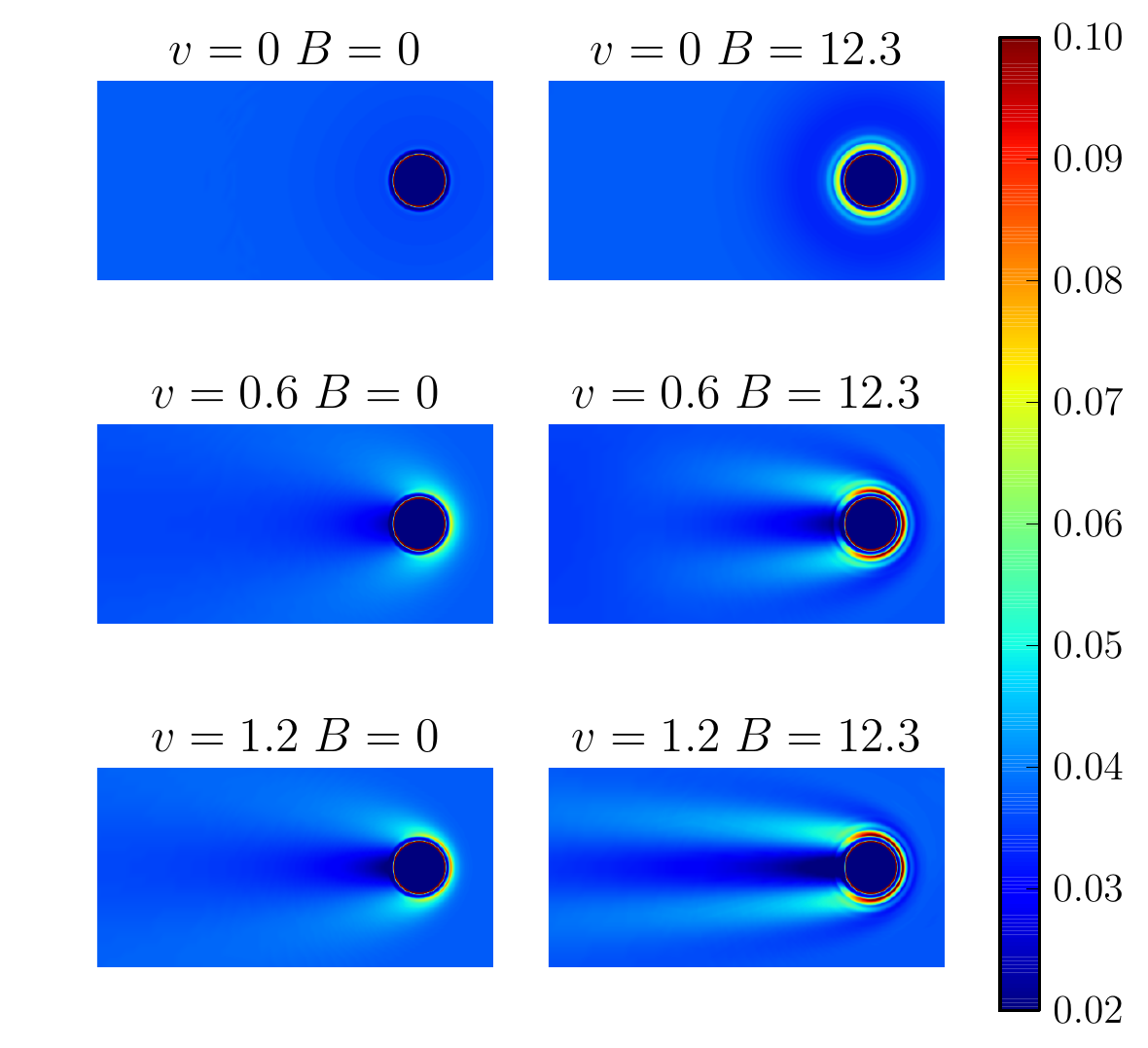}
\end{center}
\caption{\label{figure6}Density profiles for an attractive tracer, $\phi=0.117$. 
Left column: pure hard disks ($B=0$). Right column: square well attraction ($B=12.3$). 
The top right panel shows an equilibrium wetting layer, which evolves into an extended wetting 
trail as the velocity is increased.   
}
\end{figure}

In Fig.\ref{figure5} we show equilibrium radial density profiles around a static attractive tracer for $B=12.3$ 
and various volume fractions up to the numerically determined binodal. The thermodynamic path taken is indicated 
in the inset to Fig.\ref{figure1}. 
Despite the efficient algorithms we employ, our finite element DDFT solver is subject to some practical limitations 
of accuracy.  
Some slight discrepancy thus exists between the true binodal (obtained by a high accuracy common tangent 
construction on the analytic free energy) and the location of the numerical binodal which emerges from the finite 
element code. 
The position of the numerical binodal can be estimated by slowly increasing the chemical potential (for $B>B_{\rm crit}$) 
and noting where the profile jumps to a liquid asymptote (the system fills with liquid) and then reducing the 
chemical potential to observe where the density profile jumps back down to a gas asymptote.  
This procedure is subject to hysteresis, but nevertheless enables us to locate the binodal to an accuracy of a 
few percent in volume fraction. We thus avoid computationally expensive free energy calculations.  
The profiles shown in Fig.\ref{figure5} show clear evidence for the development of a finite thickness wetting layer 
as the volume fraction is increased towards coexistence.

\subsubsection{Wetting trails}
The left column of Fig.\ref{figure6} shows the two-dimensional, steady-state density profiles around an attractive tracer 
for $B=0$ and three different values of the velocity. 
These results serve as a reference with which the density profiles closer to coexistence can be compared. 
As the velocity is increased there is a pile-up of particles at the front of the tracer, as would be intuitively 
expected, leading to a high density region which becomes more compressed as the velocity is increased. 
Due to the curvature of the surface, the accumulated particles are swept first around the tracer and then 
downstream by the flow. For velocities around unity the high density, comet-like trails left behind diffuse away 
at a distance of $~2R_{\rm t}$ from the rear tracer surface. 
There is also a slight reduction in density immediately behind the tracer surface, but this is a small effect.

The right column of Fig.\ref{figure6} shows density profiles of attractive disks ($B=12.3$) around an 
attractive tracer for three different values of the velocity. 
The thermodynamic statepoint lies close to the binodal and for $v=0$ (top right panel) a wetting layer is 
clearly visible around the tracer surface.
For finite velocities the local pressure reduction directly behind the tracer tends to suppress 
the wetting layer, as the local thermodynamic statepoint of this spatial region is effectively moved further 
away from coexistence. 
In front of the tracer the wetting layer becomes compressed, leading to a strong increase 
in density within the range of the tracer attraction and an enhanced oscillatory packing structure at larger distances.

The most interesting feature of the density profiles in Fig.\ref{figure6} are the ``wetting trails" deposited behind the particle. 
These regions of high (but not quite liquid) colloidal 
density advect away from the wetting layer and, for a given tracer velocity, are much more pronounced than for the 
$B=0$ reference system. 
For a velocity close to unity (lower right panel) the wetting trail extends to around $5R_{\rm t}$ from the tracer surface before 
diffusion drives relaxation back to equilibrium. 
The lifetime, and hence spatial extent, of the wetting trails is longer than in the corresponding pure hard disk system. 
This reflects the fact that the free energy cost of liquid-like density inhomogeneities 
is low for statepoints close to the gas side of the binodal.

\begin{figure}
\begin{center}
\hspace*{-0.4cm}\includegraphics[width=9.3cm]{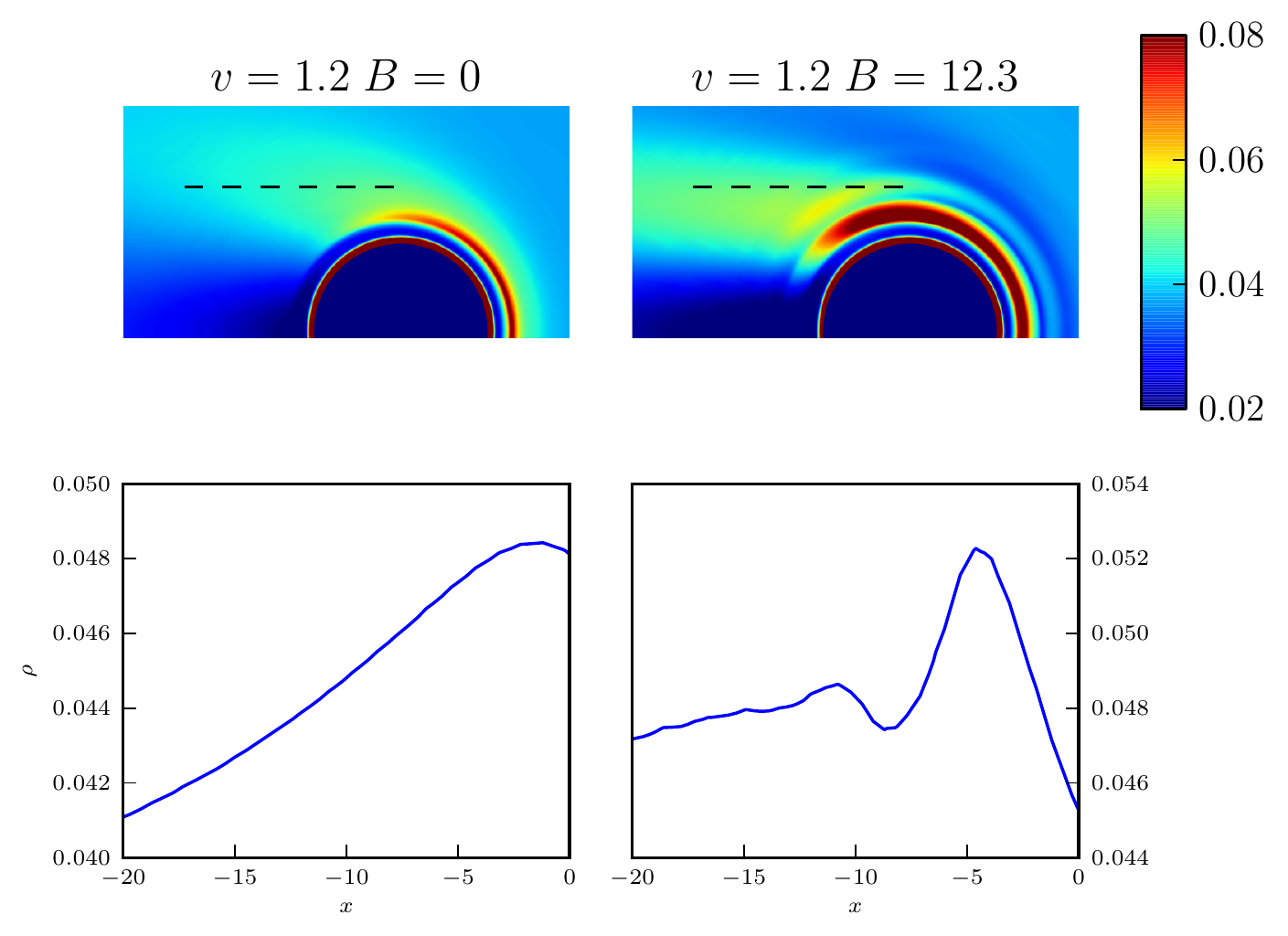}
\end{center}
\caption{\label{figure8}
Top: Focusing on the surface region for the profiles shown in the lower two panels of Fig.\ref{figure6}. 
For $B=12.3$ there exists a region of enhanced density in the upper 
left quadrant of the wetting layer. 
Bottom: slices through the two-dimensional density profiles, taken along the lines indicated in the top panels.
The flow induces additional correlation structure behind the tracer ($x<0$). 
}
\end{figure}

The genesis of the liquid trail is shown in more detail in Fig.\ref{figure8}, where we focus on the vicinity 
of the tracer surface and compare the nonequilibrium profile at $B=12.3$ with that at $B=0$ for equal volume fraction 
and tracer velocity. 
For $B=0$ (upper left panel) the contact packing peak is advected away from the tracer surface, generating a short 
trail of increased density. 
The angle at which the trail departs the surface region is approximately $20\degree$ to the negative $x$-axis.
For $B=12.3$ (upper right panel) the behaviour is quite different; here the trail originates in the 
wetting layer, rather than in the contact peak, and leaves the surface region at a departure angle of around $10\degree$. 
A nontrivial feature of this density profile is that the flow leads to region of increased density in a region 
{\it behind} the center of the tracer. 
Particles within the wetting layer are swept around the particle by the flow, where they first 
accumulate before being ejected out into the wake.  
In order to focus on this aspect of the density distribution we take a slice 
from $x=0\rightarrow -20$ (indicated by the broken line) and show the density variation in the lower two 
panels. 
For $B=0$ (lower left panel) the density in the trail decays monotonically towards the bulk value $\rho_{\rm bulk}=0.117/\pi=0.03724$. 
However, for $B=12.3$ we observe two additional peaks within the trail. 
The most prominant peak, located at $x\approx-5$, arises from the accumulation of particles swept around within the wetting layer. 
The weaker peak, at around $x\approx-11$, is not present in equilibrium and reveals nontrivial correlation effects in the wetting trail. 
Such a detailed structural description is a consequence of the high accuracy treatment of packing physics afforded by 
the fundamental measures excess free energy functional \cite{oettel}.   

\subsubsection{Friction coefficient}
In the right hand panel of Fig.\ref{figure4} we show the friction coefficient, $\xi$, as a function of velocity for 
$B=0$ and $B=12.5$. 
As previously, we are restricted to intermediate velocities by the limitations of our numerical algorithms. 
As was the case for a purely repulsive tracer (left hand panel), the friction reduces 
as a function of velocity. 
The primary reason for this reduction is again the lubrication afforded by the packing shells which build 
up at the font of the tracer.  
In contrast to pure hard disks, for which the friction depends only on the tracer contact density 
$\rho(R_{\rm t}+R)$, the frictional force of attractive disks depends also upon the density distribution over 
the entire range of the attractive potential \eqref{wet_potential}.



\section{Discussion}\label{discussion}
In this paper we have considered the active microrheology of a two-dimensional colloidal suspension at thermodynamic statepoints 
close to the gas-liquid phase boundary. Applying DDFT to study the response of the suspension to a hard tracer moving with constant 
velocity $v$ we obtain cavitating density profiles, for which a large region of low colloidal volume fraction develops behind the 
tracer. The tendency of the flow to cavitate increases as the bulk osmotic pressure approaches the vapour pressure and as the velocity 
of the tracer is increased, as described by the colloidal cavitation number ${\sf Ca}$. 
When applied to an attractive tracer at a statepoint close to the gas side of the binodal we find that the wetting of the tracer 
surface interacts in a complex way with the imposed flow field, leading to the generation of long wetting trails.
We hope that the phenomena predicted in the present study can serve to stimulate future colloidal experiments and Brownian dynamics 
simulations. 
However, there remain open issues and possibilities for future investigation. 

Throughout the present work we have neglected entirely any effects arising from solvent hydrodynamics. 
Within this ``free draining" approximation the solvent simply flows unperturbed through the tracer. 
This is clearly an idealization.  
The simplest first step beyond this approach would be to use a Stokes flow profile \cite{batchelor} around the tracer as 
input to the DDFT equation \eqref{stDDFT}. 
In many real suspensions the hydrodynamic radius of the tracer is less than the range of its potential interaction with 
the colloids. 
The tracer thus comes into direct physical contact with the colloids, rather than being kept apart by hydrodynamic 
lubrication forces. 
In such cases our theoretical calculations should capture qualitatively the correct phenomenology.

Although we have restricted our attention to fixed tracer radii and selected thermodynamic 
statepoints, a more extensive exploration of the parameter space is desirable and is currently underway. 
In particular, we expect that employing larger tracer radii will serve to enhance the phenomena identified 
in this work. 
Increasing the size of the tracer would lead to a correspondingly increased area of reduced pressure in the 
wake, which would very likely enhance the cavitation effect. Moreover, a larger attractive tracer 
has lower geometrical curvature and thus approximates more accurately a planar wall, leading to thicker wetting layers. 
We also plan to investigate the physics of multiple tracer particles for which wetting or cavitation induced 
interactions will play an important role. For example, a cavitation induced interaction between tracer particles 
could lead to steady state ordering effects in sedimentation experiments. 

We have focused on steady states, but the transient response to time-dependent driving is also a topic 
of considerable interest. 
Our methods are well suited to such studies and preliminary investigations show, for example, 
that the cavitation bubbles shown in Fig.\ref{figure2} develop on a slow timescale. 
The investigation of time-dependent response to both start-up flow (for which the stationary tracer is 
suddenly pulled with a fixed velocity) and oscillatory driving will be the subjects of future 
investigations. 
When applying the former protocol we anticipate the occurance of a transient ``overshoot" in the frictional force, 
analogous to that observed in macroscopic start-up rheological tests \cite{zausch}. 
For a recent study of transient nonlinear microrheology see \cite{zia}. 

Finally, it is important to point out that our simple mean field approach neglects fluctuations, 
which may be important in experiment and thus leaves some important open questions. 
For example, the inception of a colloidal cavity following the onset of flow probably occurs via a process of nucleation 
and growth, however the precise nature of the nucleation sites and the character of the subsequent growth remain unclear. 
Another interesting problem is whether capillary wave-type fluctuations in the interface of the cavitation bubble 
could lead to instability and ``pinch off" events, whereby a bubble of low colloidal density first 
detaches from the main cavity and then collapses further downstream.  
It is known that shearing the planar interface between colloidal liquid and gas phases tends to suppress capillary 
waves \cite{aarts_capillary}, but whether this effect is sufficient to prevent pinch-off fluctuations remains to be seen.

\begin{acknowledgments}
We thank V.~Trappe for helpful discussions. 
\end{acknowledgments}



\begin{thebibliography}{}

\bibitem{macosko}
C.W.~Macosko, {\it Rheology: Principles, Measurements, and Applications} 
(Wiley, 1994).

\bibitem{larson}
R.G.~Larson, {\it The Structure and Rheology of Complex
Fluids} (Oxford University Press, New York, 1999).

\bibitem{mewis}
J.~Mewis and N.J.~Wagner, {\it Colloidal suspension rheology} 
(Cambridge University Press, 2012).

\bibitem{joe_review}
J.M.~Brader, J.Phys.:Condens.Matter {\bf 22} 363101 (2010). 

\bibitem{butterworths}
R.G.~Larson, {\it Constitutive Equations for Polymer Melts and Solutions} 
(Butterworths, Boston, 1988).

\bibitem{swan}
J.W.~Swan and R.N.~Zia, Physics of Fluids {\bf 25} 083303 (2013) 

\bibitem{voigtmann_review}
T.~Voigtmann and M.~Fuchs, Eur.Phys.J. Special Topics, {\bf 222}, 2819 (2013).

\bibitem{squires}
T.M.~Squires, T.G.~Mason, Annu.Rev.Fluid Mech. {\bf 42}, 413 (2010).

\bibitem{seifriz}
H.~Freundlich and W.~Seifriz, Z.Phys.Chem. {\bf 104} 233 (1923).

\bibitem{bray}
A.J.~Bray, Phil.Trans.R.Soc.Lond.A {\bf 361} 781 (2003).

\bibitem{dhont2002}
J.K.G.~Dhont, in {\it Models and kinetic methods for
non-equilibrium many-body systems}, NATO science series {\bf 371} 73 (2002).

\bibitem{onuki}
A.~Onuki, {\it Phase Transition Dynamics}
(Cambridge, Cambridge University Press, 2002).

\bibitem{aarts}
D.G.A.L.~Aarts, J.H.~van der Wiel and
H.N.W.~Lekkerkerker, J.Phys.:Condens.Matt. {\bf 15} 5245 (2003)

\bibitem{derks2008}
D.~Derks, D.G.A.L.~Aarts, D.~Bonn and A.~Imhof,
J.Phys.:Condens.Matt. {\bf 20} 404208 (2008).

\bibitem{stansell}
P.~Stansell {et al.}, Phys.Rev.Lett. {\bf 96} 085701 (2006).



\bibitem{onuki1997}
A.~Onuki, J.Phys.:Condens.Matt. {\bf 9} 6119 (1997).

\bibitem{han2006}
C.C.~Han, Y.~Yao, R.~Zhang and E.K.~Hobbie,
Polymer {\bf 47} 3271 (2006).

\bibitem{saito2003}
S.~Saito {\it et al.} Macromolecules {\bf 36} 3745 (2003).

\bibitem{ackerson}
B.J.~Ackerson and P.N.~Pusey, Phys.Rev.Lett. {\bf 61} 1033 (1988)

\bibitem{besseling2012}
T.H.~Besseling et al., Soft Matter 8 2931 (2012)

\bibitem{batchelor}
G.K.~Batchelor, {\it An introduction to fluid dynamics} 
(Cambridge University Press, 2000). 

\bibitem{chen_isrealachvili}
Y.L.~Chen and J.~Isrealachvili, Science {\bf 252} 1157 (1991).

\bibitem{gardiner}
C.W.~Gardiner, {\it Handbook of Stochastic Methods}, (Springer, Berlin (1985)).

\bibitem{dhont_book}
J.~K.~G. Dhont, {\it An introduction to dynamics of colloids} (Elsevier, Amsterdam, 1996).

\bibitem{Hansen06}
J.~P.~Hansen and I.~R. McDonald, {\it Theory of Simple Liquids}, 3rd ed.
  (Academic Press, London, 2006).
  
\bibitem{evans79}
R.~Evans, Adv. Phys. {\bf 28},  143  (1979).  

\bibitem{reinhardtbrader}
J.~Reinhardt and J.M.~Brader, Phys.Rev.E {\bf 85} 011404 (2012). 

\bibitem{evans92}
R.~Evans, in {\it Fundamentals of inhomogeneous fluids} 
Edited by D.~Henderson (Dekker, New York, 1992).

\bibitem{oettel}
R.~Roth, K.~Mecke and M.~Oettel, J.Chem.Phys. {\bf 136} 181101 (2012).

\bibitem{roland_review}
R.~Roth, J-Phys.:Condens.Matter {\bf 22} 063102 (2010).



\bibitem{barker}
J.A.~Barker and D.~Henderson, 
J.Chem.Phys. {\bf 47} 4714 (1967).

\bibitem{reinhardt2013a}
J.~Reinhardt and J.M.~Brader, 
EPL {\bf 102} 28011 (2013).



\bibitem{Bangerth2007}%
  \BibitemOpen
  \bibfield{author}{%
  \bibinfo {author} {W.~Bangerth}, \bibinfo
  {author} {R.~Hartmann},\ and\ \bibinfo {author}
  {G.~Kanschat},\ }%
  \bibfield{journal}{%
  \bibinfo {journal} {ACM Trans. Math. Softw.}\ }%
  \textbf{\bibinfo {volume} {33}} (\bibinfo {year}
  {2007}).


\bibitem{Geuzaine2009}%
  \BibitemOpen
  \bibfield{author}{%
  \bibinfo {author} {C.~Geuzaine}\ and\ \bibinfo
  {author} {J.-F.\ Remacle},\ }%
  \bibfield{journal}{%
  {\bibinfo {journal} {International Journal for
  Numerical Methods in Engineering}}\ }%
  \textbf{\bibinfo {volume} {79}},\ \bibinfo {pages} {1309} (\bibinfo {year}
  {2009}).




\bibitem{abraham}
F.F.~Abraham, J.Chem.Phys. {\bf 68} 3713 (1978).

\bibitem{sullivan}
D.E.~Sullivan, D.~Levesque and J.J.~Weiss, J.Chem.Phys. {\bf 72} 1170 (1980).

\bibitem{tarazona}
P.~Tarazona and R.~Evans,
Mol.Phys. {\bf 52} 847 (1984). 

\bibitem{henderson}
J.R.~Henderson and F.~van Swol, 
Mol.Phys. {\bf 56} 1313 (1985). 

\bibitem{bayer}
M.~Bayer {\it et al.} Phys.Rev.E {\bf 76} 011508 (2007). 

\bibitem{hajnal}
D.~Hajnal, J.M.~Brader and R.~Schilling, Phys.Rev.E {\bf 80} 021503 (2009). 

\bibitem{gazuz}
I.~Gazuz, A.M.~Puertas, T.~Voigtmann and M.~Fuchs, 
Phys. Rev. Lett. {\bf 102} 248302 (2009).


\bibitem{power}
M.~Schmidt and J.~M. Brader, J. Chem. Phys. {\bf 138} 214101 (2013). 

\bibitem{noz}
J.M.~Brader and M.~Schmidt, J. Chem. Phys. {\bf 139} 104108 (2013). 

\bibitem{gelfand}
M.P.~Gelfand and R.~Lipowsky, Phys.Rev.B {\bf 36} 8725 (1987).

\bibitem{upton}
P.J.~Upton, J.O.~Indekeu and J.M.~Yeomans, Phys.Rev.B, {\bf 40} 666 (1989).

\bibitem{bieker}
T.~Bieker and S.~Dietrich, Physica A, {\bf 252} 85 (1998). 

\bibitem{streeter}
V.L.~Streeter, {\it Handbook of Fluid Dynamics} (McGraw-Hill, 1961). 

\bibitem{tarazona_evans}
P.~Tarazona and R.~Evans, Mol.Phys. {\bf 48} 799 (1983).

\bibitem{dietrich}
S.~Dietrich, {\it Wetting phenomena in phase transitions and critical phenomena} 
Vol.12, eds. C.~Domb and J.L.~Lebowitz (Academic press, London, 1988). 

\bibitem{wet1}
J.M.~Brader, R.~Evans, M.~Schmidt, and H.~Lowen,
J.Phys.Condens.Matter {\bf 14}, L1 (2002).

\bibitem{wet3}
D.G.A.L.~Aarts, R.P.A.~Dullens, D.~Bonn, R.~van Roij,
and H.N.W.~Lekkerkerker, J. Chem. Phys. {\bf 120}, 1973 (2004).

\bibitem{roth_brader_schmidt}
R.~Roth, J.M.~Brader and M.~Schmidt, EPL {\bf 63} 549 (2003).  

\bibitem{wet2}
M.~Dijkstra and R.~van Roij, Phys. Rev. Lett. {\bf 89}, 208303 (2002).

\bibitem{col_pol_review}
R.~Evans {\it et al.}, Phil.Trans. {\bf 359} 961 (2001).

\bibitem{squaregradient}
J.M.~Brader and R.~Evans, EPL {\bf 49} 678 (2000). 

\bibitem{els}
E.H.A.~de Hoog and H.N.W.~Lekkerkerker, J.Phys.Chem.B {\bf 103} 5274 (1999).





\bibitem{gil1}
T.~Gil and L.V.~Mikheev, 
Phys.Rev.E {\bf 52} 772 (1995).

\bibitem{gil2}
T.~Gil and J.H.~Ipsen, 
Phys.Rev.E {\bf 55} 1713 (1997).

\bibitem{gil3} 
T.~Gil, J.H.~Ipsen and C.F.~Tejero, 
Phys.Rev.E {\bf 57} 3123 (1998).

\bibitem{brader_kruger}
J.M.~Brader and M.~Kr\"uger, Mol.Phys. {\bf 109} 1029 (2011). 

\bibitem{zausch}
J.~Zausch {\it et al.}, J.Phys.:Condens.Matter {\bf 20} 404210 (2008). 

\bibitem{zia}
R.N.~Zia and J.F.~Brady, J.Rheol. {\bf 57} 457 (2013). 

\bibitem{aarts_capillary}
D.G.A.L.~Aarts, M.~Schmidt, and H.N.W.~Lekkerkerker, Science {\bf 304}, 847 (2004). 


\end{thebibliography}
\end{document}